\documentclass[twocolumn,prd,showpacs,showkeys,nofootinbib]{revtex4-1}
\usepackage{amsmath,amssymb,pgf,graphicx,color,url,array}
\usepackage{tabularx}
\usepackage{wasysym}
\usepackage{longtable}
\newcommand{\be}{\begin{equation}}
\newcommand{\ee}{\end{equation}}
\newcommand{\todo}[1]{{}}

\usepackage[inline]{enumitem}
\newcommand{\xmax}{$\mbox{X}_{\mbox{max}}$}

\def\itemrange#1{%
\addtocounter{enumi}{1}%
\edef\labelenumi{\theenumi--\noexpand\theenumi.}%
\addtocounter{enumi}{-1}%
\addtocounter{enumi}{#1}%
\item
\def\labelenumi{\theenumi.}}

\begin{document}
\vskip -0.9cm
\title{Mass composition of ultra-high-energy cosmic rays with the
  Telescope Array Surface Detector Data}

\author{R.U.~Abbasi$^{1}$,
M.~Abe$^{2}$,
T.~Abu-Zayyad$^{1}$,
M.~Allen$^{1}$,
R.~Azuma$^{3}$,
E.~Barcikowski$^{1}$,
J.W.~Belz$^{1}$,
D.R.~Bergman$^{1}$,
S.A.~Blake$^{1}$,
R.~Cady$^{1}$
% ...
\\~\\
{\footnotesize\it
$^{1}$ High Energy Astrophysics Institute and Department of Physics and Astronomy, University of Utah, Salt Lake City, Utah, USA \\
$^{2}$ The Graduate School of Science and Engineering, Saitama University, Saitama, Saitama, Japan \\
$^{3}$ Graduate School of Science and Engineering, Tokyo Institute of Technology, Meguro, Tokyo, Japan \\
$^{4}$ Department of Physics and The Research Institute of Natural Science, Hanyang University, Seongdong-gu, Seoul, Korea \\
$^{5}$ Department of Physics, Tokyo University of Science, Noda, Chiba, Japan \\
$^{6}$ Department of Physics, Kindai University, Higashi Osaka, Osaka, Japan \\
$^{7}$ Service de Physique ThÃ©orique, UniversitÃ© Libre de Bruxelles, Brussels, Belgium \\
$^{8}$ Institute for Cosmic Ray Research, University of Tokyo, Kashiwa, Chiba, Japan \\
$^{9}$ Graduate School of Science, Osaka City University, Osaka, Osaka, Japan \\
$^{10}$ Kavli Institute for the Physics and Mathematics of the Universe (WPI), Todai Institutes for Advanced Study, University of Tokyo, Kashiwa, Chiba, Japan \\
$^{11}$ Information Engineering Graduate School of Science and Technology, Shinshu University, Nagano, Nagano, Japan \\
$^{12}$ Faculty of Engineering, Kanagawa University, Yokohama, Kanagawa, Japan \\
$^{13}$ Interdisciplinary Graduate School of Medicine and Engineering, University of Yamanashi, Kofu, Yamanashi, Japan \\
$^{14}$ Astrophysical Big Bang Laboratory, RIKEN, Wako, Saitama, Japan \\
$^{15}$ Department of Physics, Sungkyunkwan University, Jang-an-gu, Suwon, Korea \\
$^{16}$ Department of Physics, Tokyo City University, Setagaya-ku, Tokyo, Japan \\
$^{17}$ Institute for Nuclear Research of the Russian Academy of Sciences, Moscow, Russia \\
$^{18}$ Advanced Research Institute for Science and Engineering, Waseda University, Shinjuku-ku, Tokyo, Japan \\
$^{19}$ Department of Physics, Chiba University, Chiba, Chiba, Japan \\
$^{20}$ Department of Physics, School of Natural Sciences, Ulsan National Institute of Science and Technology, UNIST-gil, Ulsan, Korea \\
$^{21}$ Department of Physics, Yonsei University, Seodaemun-gu, Seoul, Korea \\
$^{22}$ Academic Assembly School of Science and Technology Institute of Engineering, Shinshu University, Nagano, Nagano, Japan \\
$^{23}$ Faculty of Science, Kochi University, Kochi, Kochi, Japan \\
$^{24}$ Department of Physical Sciences, Ritsumeikan University, Kusatsu, Shiga, Japan \\
$^{25}$ Sternberg Astronomical Institute, Moscow M.V. Lomonosov State University, Moscow, Russia \\
$^{26}$ Department of Physics and Astronomy, Rutgers University - The State University of New Jersey, Piscataway, New Jersey, USA \\
$^{27}$ Earthquake Research Institute, University of Tokyo, Bunkyo-ku, Tokyo, Japan \\
$^{28}$ Department of Engineering Science, Faculty of Engineering, Osaka Electro-Communication University, Neyagawa-shi, Osaka, Japan \\
$^{29}$ Graduate School of Information Sciences, Hiroshima City University, Hiroshima, Hiroshima, Japan \\
$^{30}$ Institute of Particle and Nuclear Studies, KEK, Tsukuba, Ibaraki, Japan \\
$^{31}$ National Institute of Radiological Science, Chiba, Chiba, Japan \\
$^{32}$ CEICO, Institute of Physics, Czech Academy of Sciences, Prague, Czech Republic \\
$^{33}$ Department of Physics and Institute for the Early Universe, Ewha Womans University, Seodaaemun-gu, Seoul, Korea \\
$^{34}$ Department of Physics, Ehime University, Matsuyama, Ehime, Japan\\
$^{35}$ Faculty of Physics, M.V. Lomonosov Moscow State University, Moscow, Russia
}}\let\thefootnote\relax\footnote{$^{*}$ Deceased}
\let\thefootnote\relax\footnote{$^{**}$ Corresponding author, zhezher.yana@physics.msu.ru}

\begin{abstract}
The results on ultra-high-energy cosmic rays (UHECR) mass composition
obtained with the Telescope Array surface detector are presented. The
analysis employs the boosted decision tree (BDT) multivariate analysis
built upon 14 observables related to both the properties of the shower
front and the lateral distribution function. The multivariate
classifier is trained with Monte-Carlo sets of events induced by the
primary protons and iron. An average atomic mass of UHECR is presented
for energies $10^{18.0}-10^{20.0}\ \mbox{eV}$. The average atomic mass of
primary particles shows no significant energy dependence and
corresponds to $\langle \ln A \rangle = 2.0 \pm 0.1
(stat.) \pm 0.44 (syst.)$.  The result is
compared to the mass composition obtained by the Telescope Array with
\xmax\ technique along with the results of other
experiments. Possible systematic errors of the method are discussed.
\end{abstract}
\keywords{ultra-high-energy cosmic rays -- Telescope Array -- mass composition -- boosted decision trees}
\maketitle

\section{Introduction}
\label{sec:intro}

The Telescope Array (TA) experiment is the largest ultra-high-energy (UHE)
cosmic-ray experiment in the Northern hemisphere, located near Delta,
Utah, USA \cite{Tokuno}. TA is designed to register the extensive air
showers (EAS) caused by the UHE cosmic rays entering the
atmosphere. The experiment operates in hybrid mode and performs
simultaneous measurements of the particle density and timing at the
ground level with the surface detector array (SD)~\cite{TASD} and
the fluorescence light with 38 fluorescence telescopes grouped into
three fluorescence detector stations~\cite{Tokuno2}. The SD is an
array of 507 plastic scintillator detectors arranged on a square grid
with 1.2 km spacing covering an area of approximately 700
${\mbox{km}}^2$. Each detector is composed of two layers of 1.2 cm
thick extruded scintillator of the $3\ {\mbox{m}}^2$ effective area.

There is a continuous progress of the experimental techniques, which started
since the discovery of the cosmic rays more than a century
ago. Recently, the results of three independent experiments confirmed
the cut-off in the highest energy part~\cite{HiresGZK,AugerGZK,TAGZK}
of the cosmic ray energy spectrum. The latter was predicted in 1966 by
Greisen, Zatsepin and Kuzmin~\cite{g,zk}. Still, the origin of the
UHE cosmic rays remains unidentified. The mass
composition of the UHE cosmic rays at Earth is one of the measurable
quantities directly connected to the cosmic-ray acceleration mechanism
in the source and source population as well as it is related to the propagation of the UHECR. Moreover, the mass composition is the main source of uncertainty in the expected cosmogenic photon and
neutrino fluxes~\cite{Gelmini:2005wu,Aloisio:2015ega}. In the wider scope, one needs the mass composition for precision tests of the Lorentz-invariance~\cite{Saveliev} and to ensure the safety of the future 100~TeV colliders. The latter is based on the constraints on the black hole production derived from the stability of dense astrophysical objects, such as white dwarfs and neutron stars, which interact with the cosmic rays. Black hole production rate depends on the the energy per nucleon and thus on the mass composition of the UHECR~\cite{Sokolov:2016lba}.

The most established method for the UHECR composition analysis is
based on the measurements of the longitudinal shape of the EAS with
the fluorescence telescope. This method uses the depth of the shower
maximum \xmax\ as a composition-sensitive
observable~\cite{Gaisser:1993ix}. There are UHE composition results
available based on \xmax\ measured by the three experiments:
HiRes, Pierre Auger Observatory and Telescope Array~\cite{Abbasi2,Aab:2014aea,Hanlon}. The two
latter results are compatible within the systematic errors in \xmax\
measurement which are of the order of
$10-20~\mbox{g}/\mbox{cm}^2$ in the energy range up to $10^{19}~\mbox{eV}$~\cite{TA_AUGER_Composition_WG}.

This Paper is dedicated to an alternative approach to measure
the mass composition. The method uses solely the data of the
surface detector which has an undoubted advantage of the longer than
$95\%$ duty cycle~\cite{TASD}. Still, there is no single
observable known that has a comparable to \xmax\ sensitivity to the
mass composition, although measurements based on the risetime~\cite{Aab:2016enk,Aab:2017cgk} have come close. In this Paper we use the multivariate boosted decision tree (BDT)~\cite{Breiman,Schapire} technique based on a number
of composition-sensitive variables obtained during the reconstruction
of the SD events. The BDT method has proved itself reliable with a
number of successful applications for the astroparticle physics
experiments, see e.g.~\cite{Krause,Aab,Abbasi}.

The general scheme of the analysis is the following. The
proton-induced and iron-induced Monte-Carlo events are simulated using
the real-time calibration of the Telescope Array. The Monte-Carlo
events are stored in the same format as the SD data and are split into
three parts used in the following stages. First, a BDT classifier is
trained using the first part of the proton-induced Monte-Carlo (MC) events as a background and iron-induced events as signal. Second, the distribution of the classifier output $\xi$
for data is compared to the second part of the proton and iron-induced
MC events. The comparison results in the average atomic mass $\langle
\ln A \rangle$ of the primary particle as a function of
energy. Finally, the third part of the MC is used to estimate the bias
of the method and to introduce a correction to $\langle \ln A \rangle$ in
order to compensate it.

The Paper is organized as follows: in the Section~\ref{sec:data}
data and Monte-Carlo sets are described. Section~\ref{sec:method} is
dedicated to multivariate analysis method and its implementation to mass determination. Finally, results and discussion of the systematic uncertainties are provided in Section~\ref{sec:results}.

\section{Data set and simulations}
\label{sec:data}

\subsection{Surface detector data}
\label{subsec:sddata}

The data of the 9 years of the Telescope Array surface detector
operation from May 11, 2008 to May 10, 2017 are used in this Paper. Each event is a set
of the time-dependent signals (waveforms) from both upper and lower
layers of each triggered station. The waveforms are recorded by the
12-bit flash analog-to-digital converters (FADC) with the 50 MHz sampling rate and are converted to
MIPs~\cite{TASD} at the calibration stage. The station is marked as
saturated at this stage if the saturation effects are
significant. In the case of saturated detectors only the signal incidence time is
used in the analysis.

\subsection{Event reconstruction and cuts}
\label{subsec:reconstruction}

Surface detector array event reconstruction is done in two
steps~\cite{TAGZK}. At the first step, event geometry is reconstructed
using the time of the arrival of the shower front particles measured
by the triggered ($>0.3\ \mbox{MIP}$) counters. Shower front is approximated with empirical
functions proposed by Linsley~\cite{Linsley} and later modified in
AGASA experiment~\cite{Teshima}. Secondly, pulse heights in the
counters together with the event geometry information are used for
determining the normalization of the shower lateral distribution
profile $S_{800}$~\cite{Takeda}.

In order to determine the Linsley front curvature parameter an
additional joint fit of shower front and lateral distribution function
(LDF) is performed with 7 free parameters: $x_{\mbox{core}}$,
$y_{\mbox{core}}$, $\theta$, $\phi$, $S_{800}$, $t_0$, $a$~\cite{TAgammalim}:

\begin{equation}
t_0 \left(r \right) = t_0 + t_{plane} + a \times \left(1+r/R_L\right)^{1.5} LDF\left(r \right)^{-0.5},
\end{equation}
\begin{equation}
S \left(r \right) = S_{800} \times LDF\left(r \right),
\end{equation}
\begin{equation}
LDF\left(r \right) = f\left(r \right)/f\left(800~\mbox{m}\right),
\end{equation}
\begin{equation}
f\left(r \right) = \left(\frac{r}{R_m}\right)^{-1.2}\left(1+\frac{r}{R_m}\right)^{-(\eta-1.2)}\left(1+\frac{r^2}{R_1^2}\right)^{-0.6},
\end{equation}
\begin{equation*}
R_m = 90.0~\mbox{m},~R_1 = 1000~\mbox{m},~R_L = 30~\mbox{m},
\end{equation*}
\begin{equation*}
\eta = 3.97 - 1.79 \left(\sec\left(\theta\right) -1\right),
\end{equation*}
\begin{equation*}
r = \sqrt{{\left(x_{\mbox{core}} - x \right)}^2 + {\left(y_{\mbox{core}} - y \right)}^2},
\end{equation*}

\noindent where $x_{\mbox{core}}$, $y_{\mbox{core}}$, $x$ and $y$ are obtained from the pre-defined coordinate system of the array centered at the Central Laser Facility (CLF)~\cite{Takahashi:2011zzd}, $t_{plane}$ is the delay of the shower plane and $a$ is the Linsley front curvature parameter. Including the Linsley front
curvature, 14 composition-sensitive parameters are estimated for each
event, see Appendix A for details.

The parameters may be qualitatively split into three groups. The first group of parameters is related to the LDF which is known to be sensitive to \xmax. These are the $S_b$ for $b=3$ and $b=4.5$~\cite{Ros}, the sum of the signals of all the detectors of the event, the number of the detectors hit and ${\chi}^2 / d.o.f.$ of the LDF fit.

The second group is related to the shower front which is in turn
sensitive to both \xmax\ and the muon content of the shower. The
Linsley curvature parameter designates the shower front curvature,
while the area-over-peak of the signal, its slope and the number of detectors excluded from the fit correlate with the shower front width.

The latter group indicates the muon content of the shower. Muons cause the single peaks in FADC traces as they propagate rectilinearly and have small dispersion of arrival time. Moreover, muons induce identical signals in the upper and in the lower layers of the detector. Hence, the total number of peaks within all FADC traces, number of peaks in the detector with the largest signal, number of peaks present in the upper layer and not in the lower and vice versa, and also the asymmetry of the signal at the upper and at the lower layers of the detector are affected by the muonic component of the shower.

The following cuts are used to ensure the quality of reconstruction:

\begin{enumerate}
\item event includes 7 or more triggered stations;
\item zenith angle is below $45^{\circ}$;
\item reconstructed core position inside the array with
  the distance of at least $1200\ \mbox{m}$ from the edge of the array;
\item ${\chi}^2 / d.o.f.$ doesn't exceed 4 for both the geometry and
  the LDF fits;
\item ${\chi}^2 / d.o.f.$ doesn't exceed 5 for the joint geometry and
  LDF fit.
\item an arrival direction is reconstructed with accuracy less than $5^{\circ}$;
\item fractional uncertainty of the $S_{800}$ is less than 25 \%.
\end{enumerate}

The same cuts are applied to both the data and the Monte-Carlo
sets. The cuts listed above are tighter compared to the standard
analysis cuts~\cite{TAGZK} due to the additional requirement of the
curvature parameter reconstruction quality. Namely, 7 triggered
stations is required instead of 5 and additional $\chi^2$ condition
for the joint fit is included~\cite{TAgammalim}.

After the cuts, the SD data set contains 18068 events with energy
greater than $10^{18}\ \mbox{eV}$ and less than $10^{20}\ \mbox{eV}$.

\begin{figure*}
\includegraphics[width=0.95\linewidth]{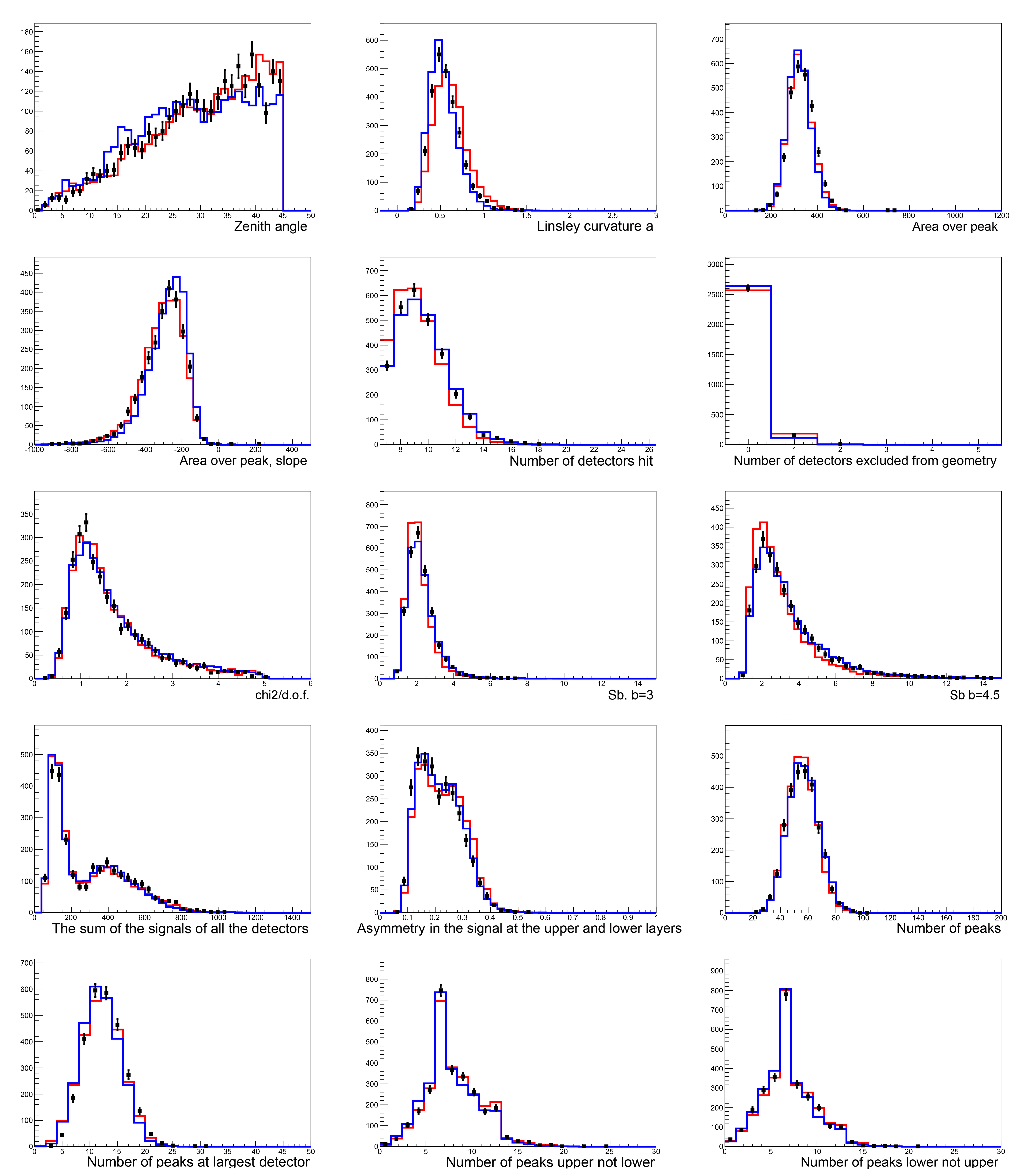}
\caption{Distributions of BDT parameters for energy bin $\log_{10} E = 18.8-19.0$. Proton MC is shown with red lines, iron MC is shown with blue lines and black dots represent the data.}
\label{histoparam}
\end{figure*}

BDT parameters distribution histograms for energy bin $\log_{10} E = 18.8 -19.0$ are denoted in Fig.~\ref{histoparam}, proton MC is shown with red lines, iron MC is shown with blue lines and black dots represent the data.

Let us discuss a contribution of individual parameters to overall BDT result. The TMVA package provides a relative importance value for each variable. The importance values are somewhat different in each energy range. Typically, the most discriminating variables are shower front curvature, $\chi^2$ and energy with importance about $8\%$. The least discriminative variables are number of detectors hit and number of detectors excluded from geometry fit with importance about $3\%$ and $1\%$ correspondingly. The remaining 11 parameters have importance value between $5\%$ and $7\%$.

\begin{figure*}
\includegraphics[width=0.65\linewidth]{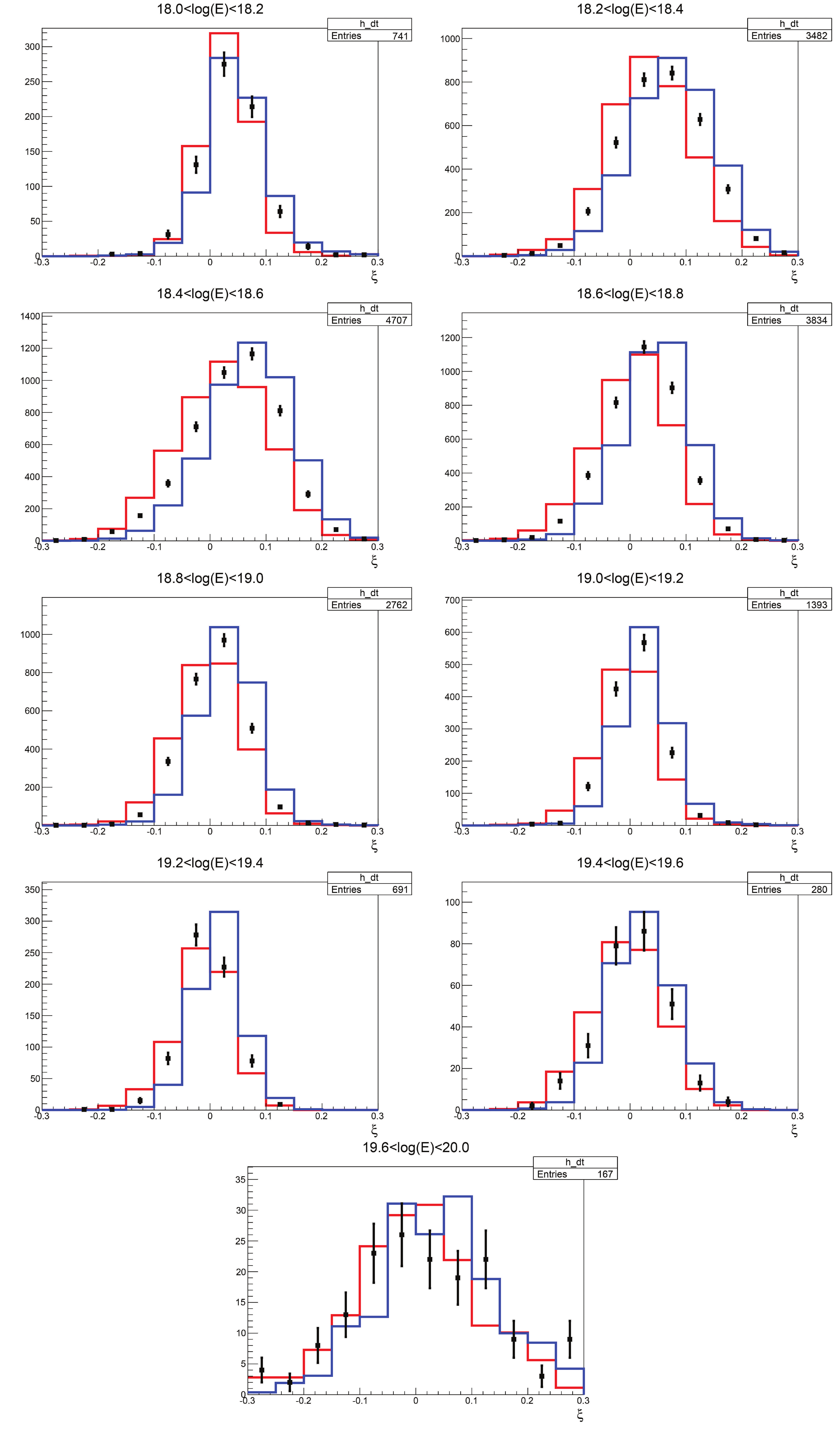}
\caption{$\xi$ parameter distribution for different energy bins. Proton MC is shown with red lines, iron MC is shown with blue lines and black dots represent the data.}
\label{xi}
\end{figure*}

\subsection{Simulations}
\label{subsec:simulations}

For the Monte-Carlo simulations, CORSIKA software package~\cite{Heck}
is used along with the QGSJETII-03 model for high-energy hadronic
interactions~\cite{Ostapchenko}, FLUKA~\cite{FLUKA,FLUKA2} for low energy hadronic
interaction and EGS4~\cite{EGS4} for electromagnetic processes.

Due to the large number of particles born in an extensive air shower,
modern computer resources available make it impractical to track
every single one in a simulation. Instead, a \textit{thinning}
procedure was proposed~\cite{Hillas}. Within thinning, all particles
with energies greater than a certain fraction of the primary energy
$\epsilon_{th}$ are followed in detail, but below the threshold only
one particle out of the secondaries produced in a certain interaction
is randomly selected. This effective particle is assigned a weight to
ensure energy conservation. The thinning level of
$\epsilon_{th}=10^{-6}$ with an additional weight limitation according
to~\cite{Kobal:2001jx} is used for simulations. The thinning allows
to achieve CPU-time efficiency, but at the same time
introduces artificial statistical
fluctuations~\cite{Gorbunov:2007vj}. The \textit{dethinning} procedure
is developed and implemented~\cite{Stokes} in order to restore the
statistical properties of the shower. The detector response is
simulated by the GEANT4 package~\cite{Agostinelli}. Real-time array
status and detector calibration information for 9 years of
observations are used for each simulated event~\cite{TAdataMC}.  Two
separate Monte-Carlo sets, for proton and iron primaries, are simulated
and stored in the same data format as the SD data. In the energy range $10^{17.5} - 10^{20.5}$ eV a set of 9800 CORSIKA showers was created. Using these showers, 200 million events were thrown on the detector for each MC set. The procedure of the Monte-Carlo set production for the Telescope Array is described in details in~\cite{AbuZayyad:2012ru}.

For each of the fourteen variables, its data and MC distributions were verified to be in the reasonable agreement. Within errors, all distributions of variables of data events lie between the proton and iron distributions.

\section{Method}
\label{sec:method}

\subsection{BDT classifier}
\label{subsec:BDT}

A number of composition-sensitive observables may be extracted from the data, and
therefore one may benefit from using the multivariate analysis
techniques. In this Paper, Boosted Decision Trees (BDT)
technique is implemented, available as a part of the ROOT Toolkit for
Multivariate Data Analysis (TMVA) package~\cite{Hocker}. The adaptive
boosting (AdaBoost) algorithm is employed~\cite{Schapire,Freund} with
the number of trees $\mbox{NTrees=1000}$.

The proton and iron Monte-Carlo sets are split into 3 parts with equal
statistics. The first part is used to build and train the BDT
classifier based on 16 variables, including zenith angle, energy and
14 composition-sensitive parameters listed in
Appendix~A. Proton-induced MC showers are used as a background and
iron-induced ones as a signal events. A separate classifier is
constructed for each energy bin with the width of $\log_{10} E = 0.2$: last two bins were merged together due to low number of data events. The
classifier is applied to the data set as well as to the two remaining
parts of the Monte-Carlo sets.

The result of the BDT classifier is a single value $\xi$ for each data
and Monte-Carlo event.  $\xi$ resides in the range $\xi \in [-1;1]$,
where $\xi = 1$ is a pure signal event , $\xi = -1$ -- pure background
event. The variable $\xi$ is used in the following one-dimensional analysis. Figure~\ref{xi} shows $\xi$ parameter distribution histograms for all the energy bins, proton MC is shown with red lines, iron MC is shown with blue lines and black dots represent the data.

\subsection{Estimation of an average atomic mass}
\label{subsec:xitolna}

Following the two-component approximation, the binned template
fitting procedure is applied to $p$, $Fe$ and data $\xi$ distributions
separately in each energy bin. The implemented method is \url{TFractionFitter} ROOT package
\cite{ROOT,TFractionFitter}. The second part of the Monte-Carlo is
used in this step to obtain the fraction of proton and iron in the
data, $\epsilon_p$ and $\epsilon_{Fe}=1-\epsilon_p$, respectively.

The first estimate of an average atomic mass is based on the
derived fraction of protons $\epsilon_p$:
\begin{equation}
\langle \ln A \rangle^{(1)} = \epsilon_p \times \ln \left( M_p \right) + (1 -
  \epsilon_p) \times \ln \left( M_{Fe} \right)\,, \label{firstestim}
\end{equation}
where $M_p = 1.0$ and $M_{Fe} = 56.0$ are average atomic masses of proton and iron nuclei.

We note that the number of proton and iron-induced simulated showers
is the same, while the trigger and reconstruction efficiency
differ. The proton fraction $\epsilon_p$ is defined as the fraction of
proton {\it simulated} events in the mixture which corresponds to the
hypothesis that $\langle \ln A \rangle^{(1)}$ is the average atomic mass of the
particles arriving to the atmosphere. It is assumed that the detector efficiency affects
the statistics of the proton and iron MC showers in the same way it
affects the proton and iron-induced events in the data.

\subsection{Bias correction}

\begin{figure}
\includegraphics[width=0.95\columnwidth]{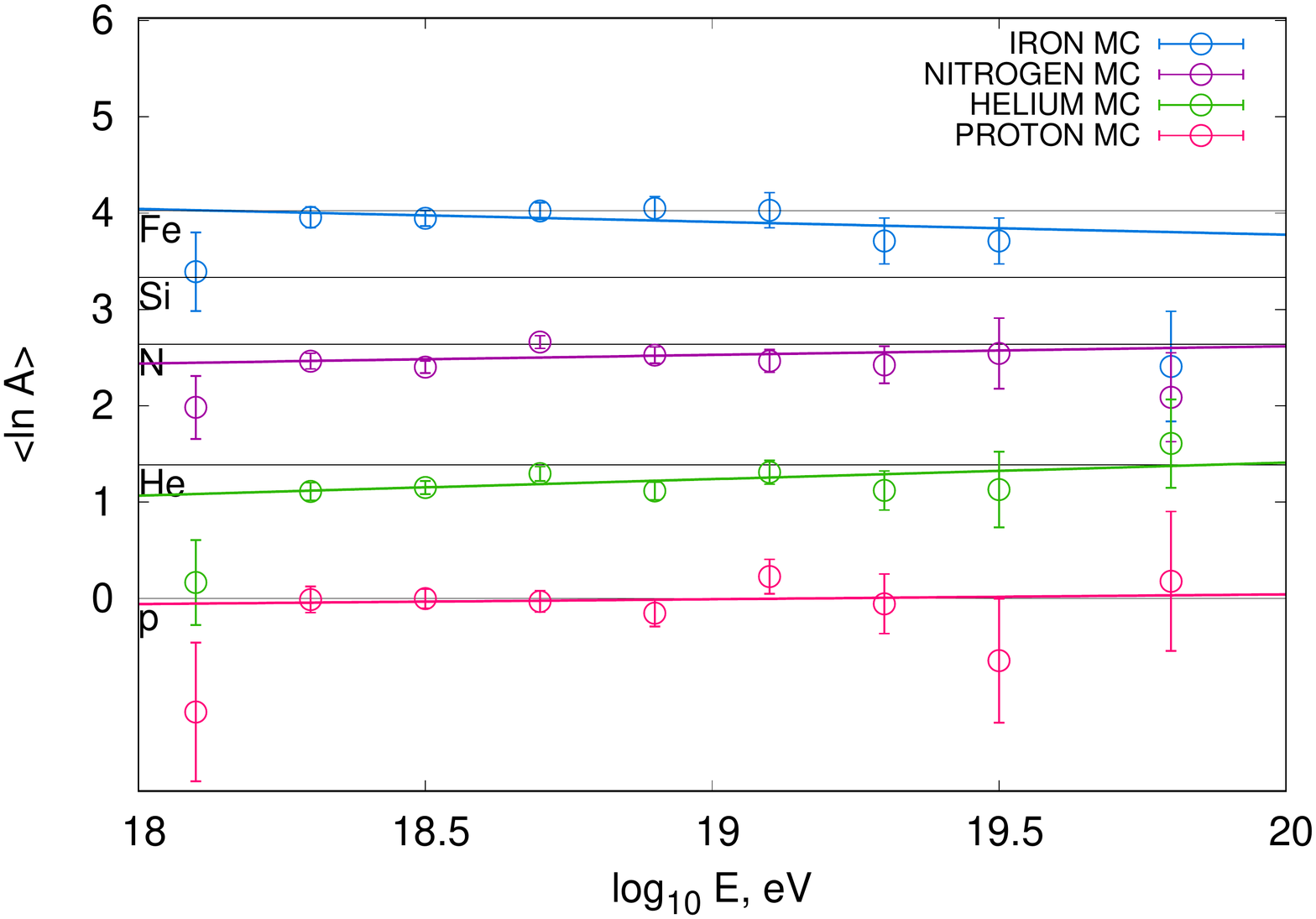}
\caption{$\langle \ln A \rangle$ approximated with a straight line for proton (red), helium (green), nitrogen (purple) and iron (blue) Monte-Carlo sets. Error bars for each $\langle \ln A \rangle$ point represent the statistical uncertainty of the method.}
\label{rails}
\end{figure}

One may go further and build the bias correction procedure based on
the Fig.~\ref{rails}. Assuming that the cosmic ray flux is composed of
particles of single type in each energy bin, it is possible to construct
the quadratic polynomial function $\ln A_{true} \left( \langle \ln A
\rangle \right)$ based on $\langle \ln A \rangle$ obtained for four MC
sets, for which the $\ln A$ values are known.

In the Figure~\ref{nonlinearcorr} uncorrected $\langle \ln A
\rangle^{(1)}$ and $\langle \ln A \rangle_{non-linear}$ obtained with
non-linear bias corrections are shown in comparison.

\section{Results and discussion}
\label{sec:results}

\subsection{Estimation of the systematic error}
\label{subsec:validation}

\begin{figure}
\includegraphics[width=0.95\columnwidth]{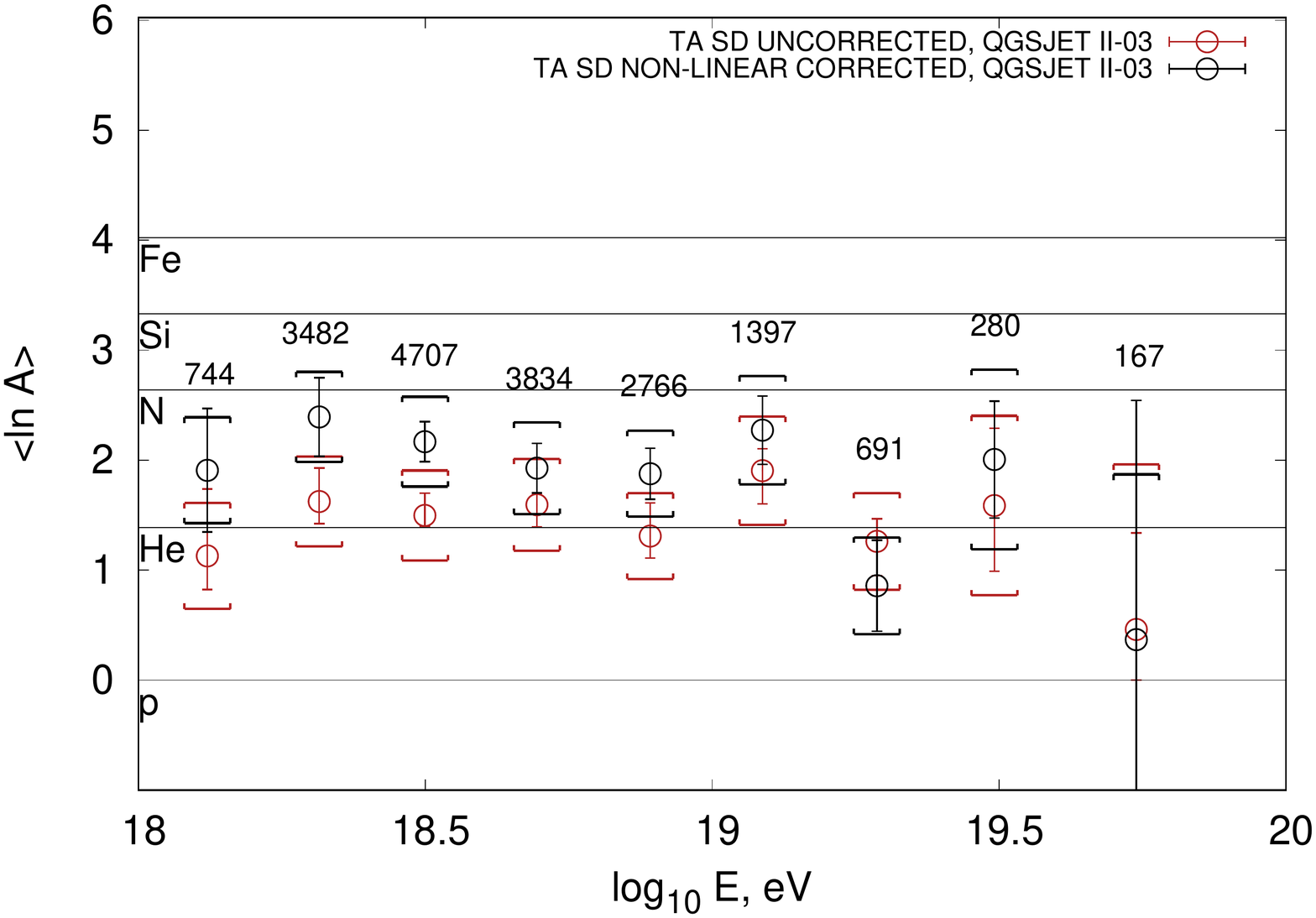}
\caption{Uncorrected $\langle \ln A \rangle^{(1)}$ in comparison with $\langle \ln A \rangle_{non-linear}$ corrected by non-linear function in each energy bin; statistical error is shown with error bars and systematic error as estimated in Section~\ref{subsec:validation} is shown with brackets of the corresponding color. Numbers represent the number of data events in the corresponding energy bin.}
\label{nonlinearcorr}
\end{figure}

\begin{figure}
\includegraphics[width=0.95\columnwidth]{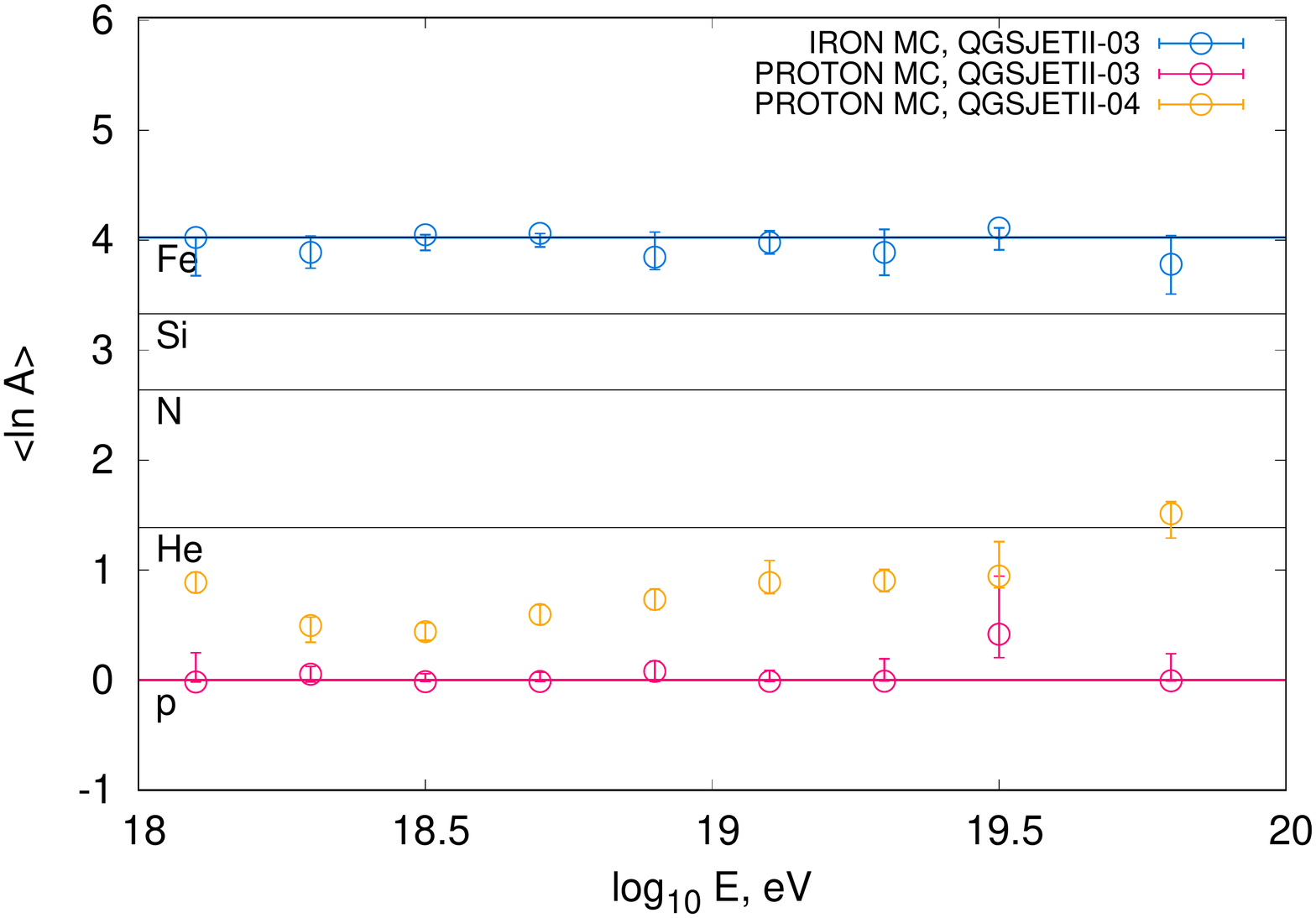}
\caption{$\langle \ln A \rangle$ approximated with a straight line for proton (red) and iron (blue) Monte-Carlo sets created with QGSJETII-03 hadronic interaction set and for proton MC set, created with QGSJETII-04 (orange line). Error bars for each $\langle \ln A \rangle$ point represent the statistical bias of the method.}
\label{qgsjet04}
\end{figure}

\begin{figure}
\includegraphics[width=0.95\columnwidth]{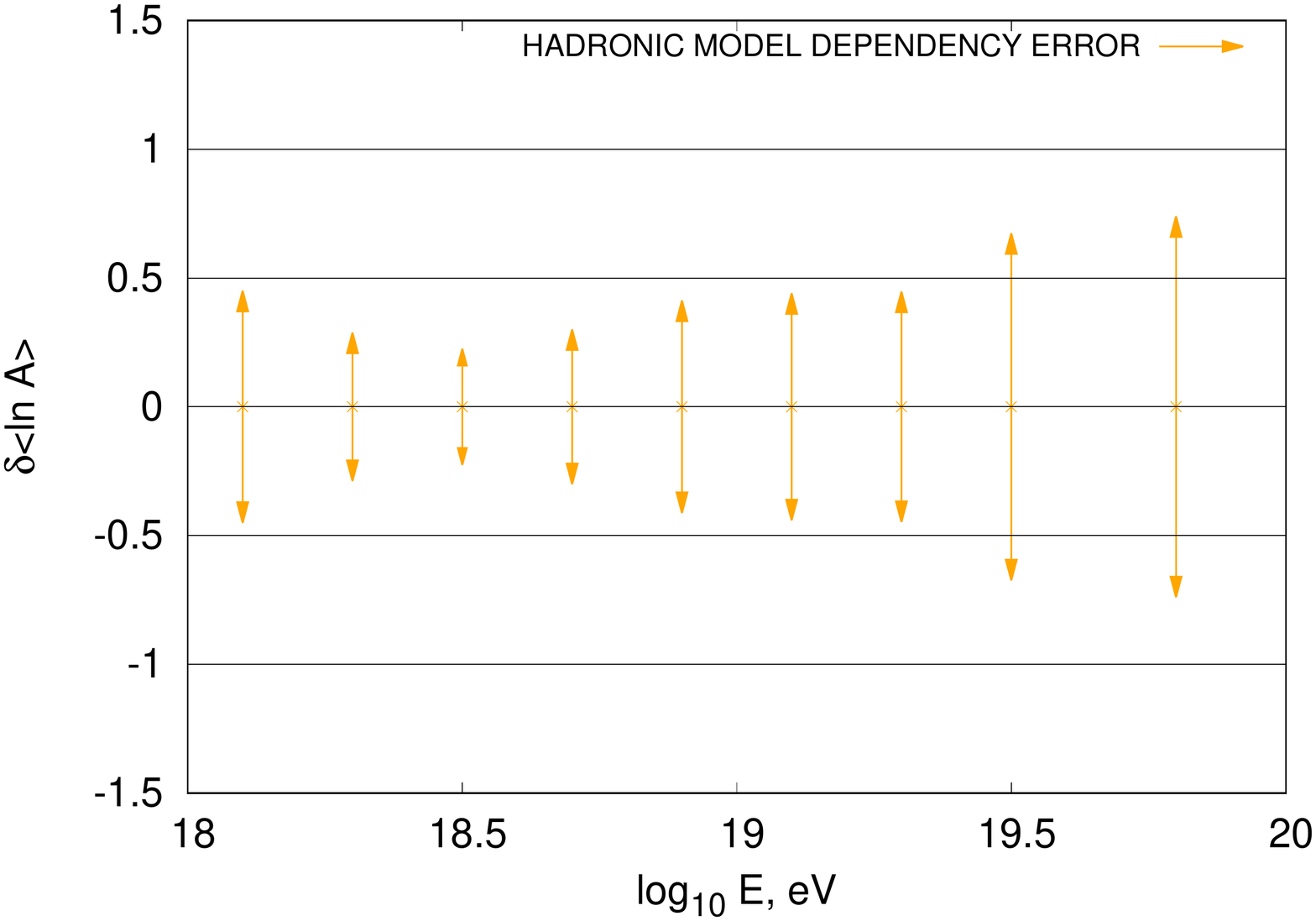}
\caption{Hadronic model dependency error of the method as a function of energy, based on a comparison with QGSJETII-04 hadronic interaction model.}
\label{deltalnAhad}
\end{figure}

\begin{figure*}
\includegraphics[width=0.65\linewidth]{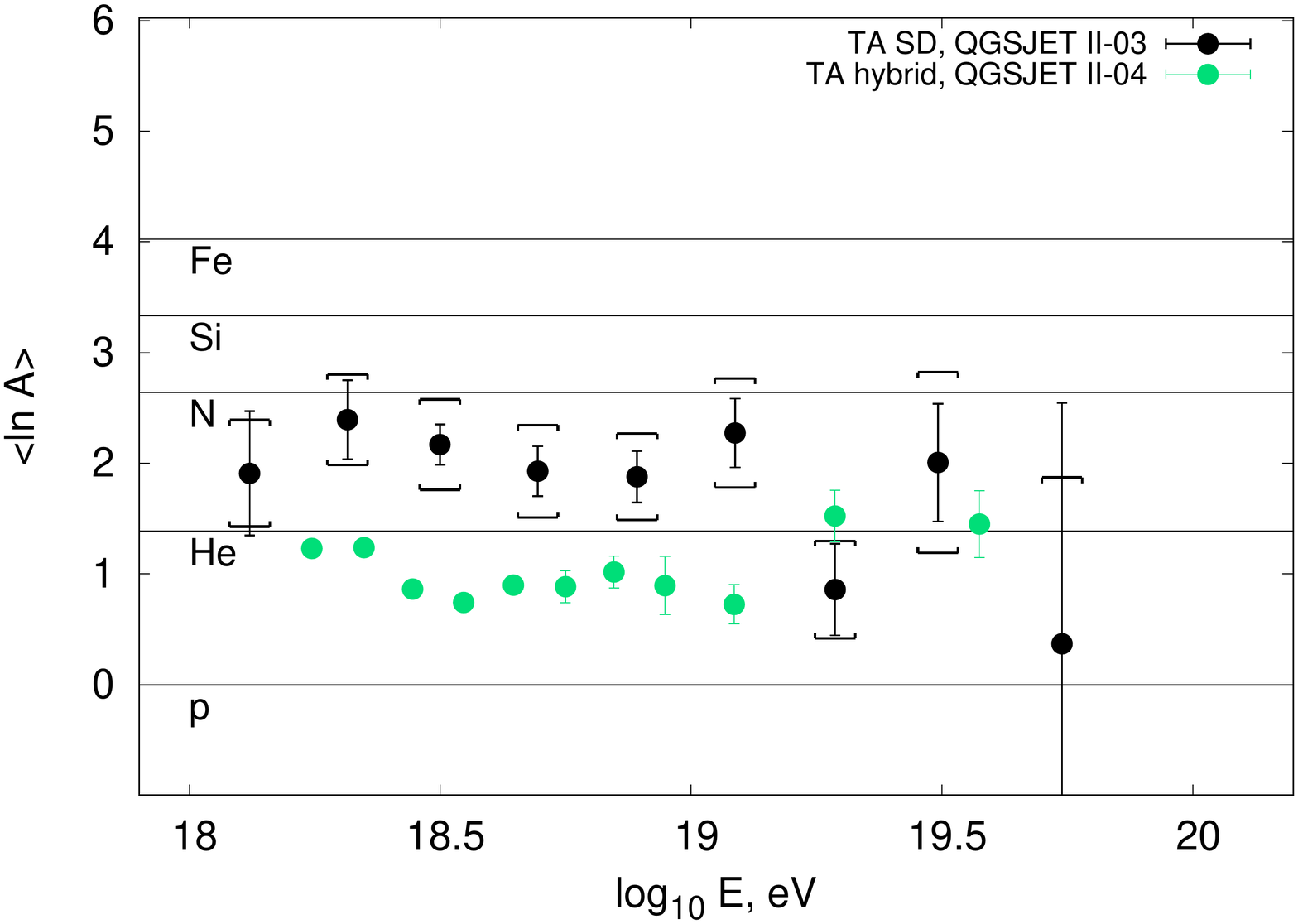}
\caption{Average atomic mass $\langle \ln A \rangle$ in comparison with the Telescope Array hybrid results \cite{Hanlon}; statistical error is shown with error bars, systematic error is shown with brackets.}
\label{Hybrid}
\end{figure*}

\begin{figure}
\includegraphics[width=0.95\columnwidth]{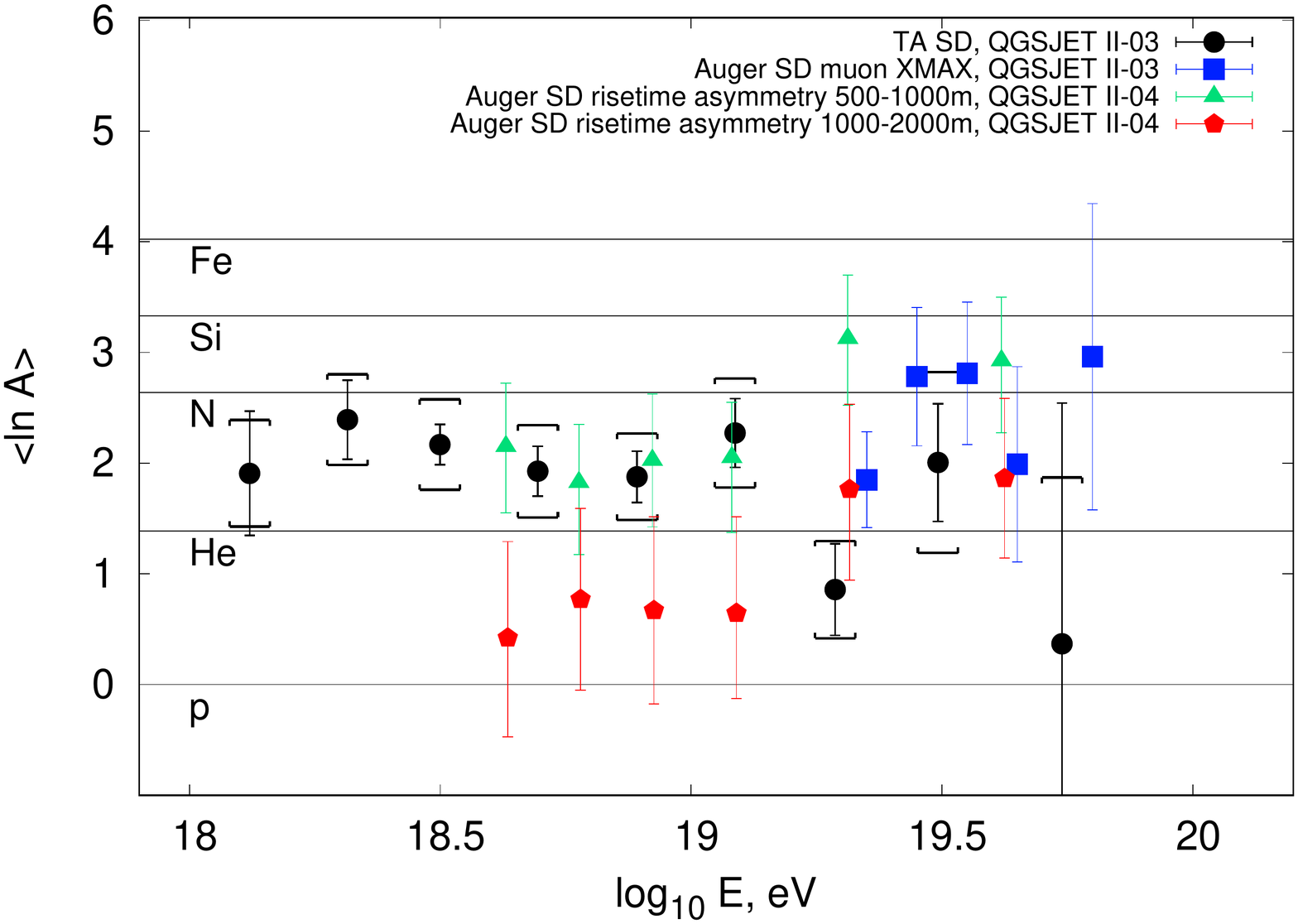}
\caption{Average atomic mass $\langle \ln A \rangle$ in comparison with the Pierre Auger Observatory $X^{\mu}_{MAX}$ and risetime asymmetry results \cite{Aab:2016enk,PierreAuger}; statistical error is shown with error bars, systematic error is shown with brackets.}
\label{Auger}
\end{figure}

\begin{figure}
\includegraphics[width=0.95\columnwidth]{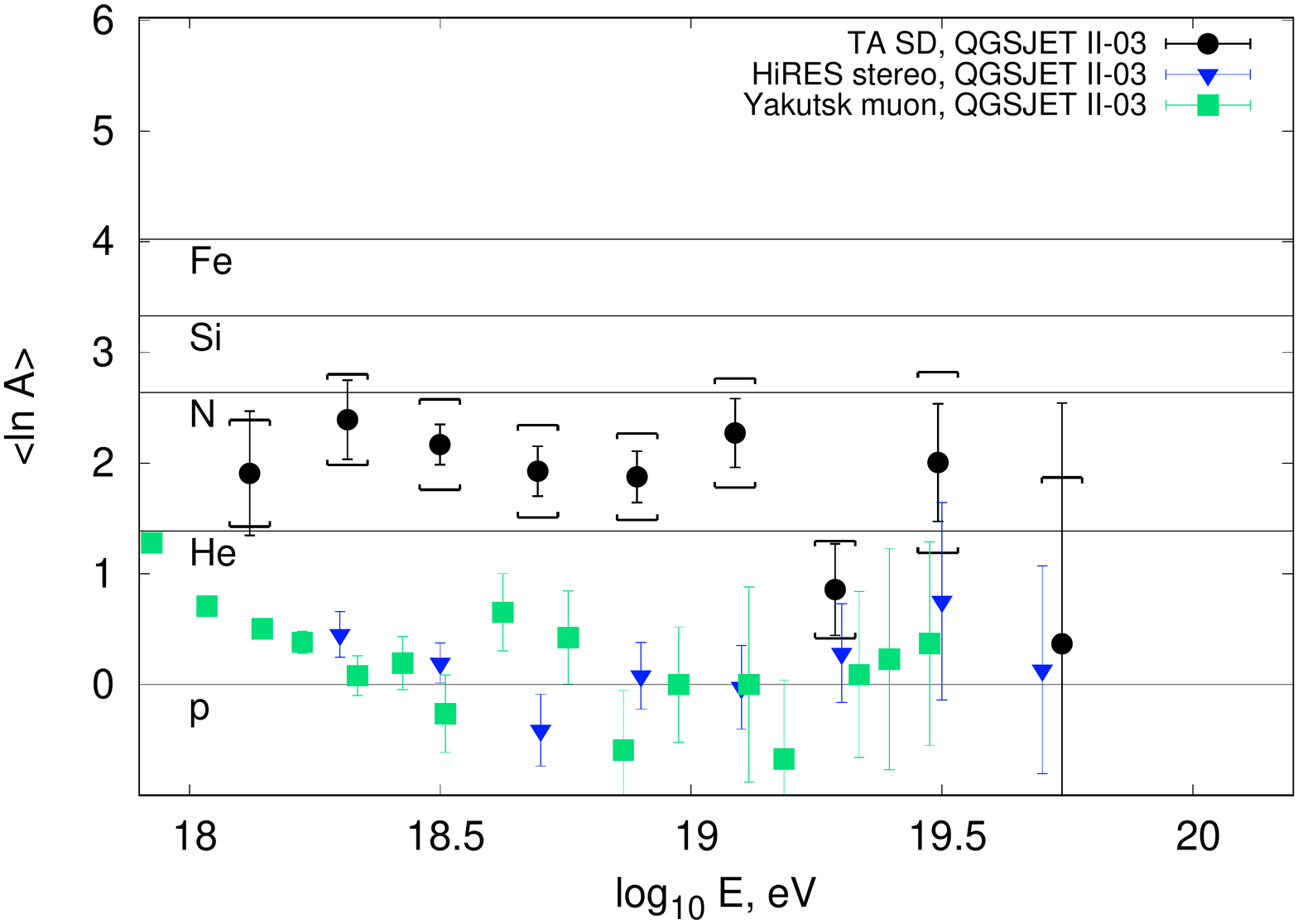}
\caption{Average atomic mass $\langle \ln A \rangle$ in comparison with the HiRes stereo results \cite{Abbasi2} and with the Yakutsk $\rho_{\mu}$ results \cite{Dedenko}; statistical error is shown with error bars, systematic error is shown with brackets.}
\label{HiRes}
\end{figure}

The non-linear correction applied for the method is based on the assumption that the obtained composition is monotype. Thus the main source for the systematic error of the method is the inability to distinguish the mixture of a given elements and the single-type-particle composition.

To derive the systematic uncertainty, in each energy bin 100 mixtures of $p$, $He$, $N$ and $Fe$ Monte-Carlo sets were created, among which 50 mixtures are random monotype, 25 are random two-component and 25 are random four-component. Its $\langle \ln A \rangle$ values were estimated with the use of \url{TFractionFitter} template fitting method and non-linear bias corrections applied and compared with the ``true'' values calculated from the known fractions. Mean systematic error is estimated as:

\begin{equation}\label{systmixed}
\delta \ln A_{syst.} = 0.44
\end{equation}

\subsection{Hadronic models dependency}
\label{subsec:hadmodels}

\todo{add models} Composition results, both derived from surface
detectors and in a hybrid mode, have a strong dependence on hadronic
models used during Monte-Carlo simulations. Besides the one used in
the above analysis, QGSJETII-04~\cite{Ostapchenko:2010vb}, an
improvement of QGSJETII-03 model, EPOS-LHC~\cite{Pierog} and
SYBILL~\cite{Fletcher} models are also widely used.

All of the hadronic interaction models are based on the collider data
and extrapolated to the UHECR energies. The analysis by the Pierre
Auger Observatory has shown the inconsistency between muon signal
predicted by simulations and data~\cite{Aab:2016hkv}. The same
conclusions were also made based on the Telescope Array SD
data~\cite{Takeishi}. This discrepancy may be the source of additional
systematic bias which may affect the observables used for the
composition study.

We study the systematic error introduced by the limited knowledge of
the hadronic interaction models based on the comparison of the two
models: QGSJETII-03 and QGSJETII-04~\cite{Ostapchenko:2010vb}. For the
latter, an additional proton Monte-Carlo set with the use of
QGSJETII-04 model is simulated. The set is subjected to the same
multivariate analysis procedure trained with the original QGSJETII-03
Monte-Carlo. The result is shown in the Fig.~\ref{qgsjet04}, while the hadronic model uncertainty as a function of energy is shown in Fig.~\ref{deltalnAhad}. The uncertainty from hadronic interaction
models is minimal at $10^{18.5}\ \mbox{eV}$ with $\delta \ln A_{hadr.} = 0.23$ and maximal at $10^{19.75}\ \mbox{eV}$ with $\delta \ln A_{hadr.} = 0.74$.

\subsection{Composition}
\label{subsec:composition}

Mean logarithm of atomic mass as a function of energy without bias
corrections and with the linear corrections applied is shown in Fig.~\ref{nonlinearcorr}. Within the errors, the
average atomic mass of primary particles shows no significant energy
dependence and corresponds to $\langle \ln A \rangle = 2.0 \pm 0.1
(stat.) \pm 0.44 (syst.)$.

TA SD composition results in comparison with TA hybrid results are
shown in Fig. \ref{Hybrid}. Comparisons with Pierre Auger Observatory
SD $X^{\mu}_{MAX}$ based on muon density and muon arrival times and azimuthal risetime asymmetry, HiRes stereo \xmax\ and
Yakutsk muon detector results are shown in
Fig. \ref{Auger} and \ref{HiRes}, respectively. We mention that while there exist composition results based on the Pierre Auger Observatory hybrid observations~\cite{Aab:2017njo}, we focus only on the comparison with the corresponding surface detector results. The obtained
composition is qualitatively consistent with the TA hybrid and the
Pierre Auger Observatory results, while all the points lie higher than the pure proton composition observed by HiRes and Yakutsk.

\section*{Acknowledgment}
\label{sec:acknowledgment}

The Telescope Array experiment is supported by the Japan Society for the Promotion of Science(JSPS) through Grants-in-Aid for Priority Area 431, for Specially Promoted Research JP21000002, for Scientific Research (S) JP19104006, for Specially Promote Research JP15H05693, for Scientific Research (S) JP15H05741 and for Young Scientists (A) JPH26707011; by the joint research program of the Institute for Cosmic Ray Research (ICRR), The University of Tokyo; by the U.S. National Science Foundation awards PHY-0601915, PHY-1404495, PHY-1404502, and PHY-1607727; by the National Research Foundation of Korea (2017K1A4A3015188 ; 2016R1A2B4014967; 2017R1A2A1A05071429, 2016R1A5A1013277);
by IISN project No. 4.4502.13, and Belgian Science Policy under IUAP VII/37 (ULB). The development and application of the multivariate
analysis method is supported by the Russian Science Foundation grant
No. 17-72-20291 (INR).
The foundations of Dr. Ezekiel R. and Edna Wattis Dumke, Willard L. Eccles, and George S. and Dolores Dore Eccles all helped with generous donations. The State of Utah supported the project through its Economic Development Board, and the University of Utah through the Office of the Vice President for Research. The experimental site became available through the cooperation of the Utah School and Institutional Trust Lands Administration (SITLA), U.S. Bureau of Land Management (BLM), and the U.S. Air Force.  We appreciate the assistance of the State of Utah and Fillmore offices of the BLM in crafting the Plan of Development for the site. Patrick Shea assisted the collaboration with valuable advice on a variety of topics. The people and the officials of Millard County, Utah have been a source of steadfast and warm support for our work which we greatly appreciate. We are indebted to the Millard County Road Department for their efforts to maintain and clear the roads which get us to our sites. We gratefully acknowledge the contribution from the technical staffs of our home institutions. An allocation of computer time from the Center for High Performance Computing at the University of Utah is gratefully acknowledged. The cluster of the Theoretical Division of INR RAS was used for the numerical part of the work.

\section*{Appendix A: Composition-sensitive variables}
\label{app:variables}

In this work, a set of fourteen composition-sensitive variables is used:

\begin{enumerate}[series=MyList]
\item Linsley front curvature parameter, as described in section \ref{subsec:reconstruction}.
\itemrange{1} Area-over-peak (AoP) of the signal at 1200 m and AoP slope parameter \cite{Abraham}:
\end{enumerate}

\begin{changemargin}{0.75 cm}{0 cm}
Given a time resolved signal from a surface station, one may calculate
its peak value and area, which are both well-measured and not much affected by fluctuations.

$AoP\left(r \right)$ is fitted with a linear fit:

\begin{equation*}
AoP\left(r \right) = \alpha - \beta \left( r/r_0 - 1.0 \right),
\end{equation*}

where $r_0=1200$\,m, $\alpha$ is $AoP\left(r \right)$ value at 1200 m
and $\beta$ is its slope parameter.

\end{changemargin}

\begin{enumerate}[resume=MyList]
\item Number of detectors hit.
\item Number of detectors excluded from the fit of the shower front by the reconstruction procedure~\cite{AbuZayyad3}.
\item $\chi^2/d.o.f.$ of the joint geometry and LDF fit.
\itemrange{1} $S_b$ parameter for $b=3$ and $b=4.5$~\cite{Ros}. The definition of
the parameter is the following:

\begin{equation*}
S_b = \sum_{i=1}^{N} \biggl[ S_i \times {\left( \frac{r_i}{r_0} \right)}^b \biggr],
\end{equation*}

\noindent where $S_i$ is the signal of {\it i}-th detector, $r_i$ is the
distance from the shower core to this station in meters and $r_0  =
1200\ \mbox{m}$ -- reference distance. The value $b = 3$ and $b = 4.5$
are used as they provide the best separation.

\item The sum of the signals of all the detectors of the event.
\item Asymmetry of the signal at the upper and lower layers of detectors.
\item Total number of peaks within all FADC (flash analog-to-digital converter) traces.

\noindent This value is summed over both upper and lower layers of all stations of the event. To suppress
  accidental peaks resulting from FADC noise, the peak is defined as a time bin with a signal exceeding 0.2 vertical equivalent muons (VEM) with the value higher than signals of the 3 preceding and 3 consequent time bins.

\item Number of peaks for the detector with the largest signal.
\item Number of peaks present in the upper layer and not in the lower.
\item Number of peaks present in the lower layer and not in the upper.
\end{enumerate}


\begin{thebibliography}{99}

\bibitem{Tokuno}
 H.~Tokuno {\it et al.} [Telescope Array Collaboration], J.\ Phys.\ Conf.\ Ser.\  {\bf 293}, 012035 (2011).

 \bibitem{TASD}
  T.~Abu-Zayyad {\it et al.} [Telescope Array Collaboration], Nucl.\ Instrum.\ Meth.\ A {\bf 689}, 87 (2013) [arXiv:1201.4964 [astro-ph.IM]].

\bibitem{Tokuno2}
  H.~Tokuno {\it et al.} [Telescope Array Collaboration], Nucl.\ Instrum.\ Meth.\ A {\bf 676}, 54 (2012) [arXiv:1201.0002 [astro-ph.IM]].

\bibitem{HiresGZK}
R.~U.~Abbasi {\it et al.} [HiRes Collaboration], Phys.\ Rev.\ Lett.\  {\bf 100}, 101101 (2008) \& R.~U.~Abbasi {\it et al.} [HiRes Collaboration], Astropart. Phys. {\bf 32} (2010).


\bibitem{AugerGZK}
J.~Abraham {\it et al.} [Pierre Auger Collaboration],  Phys.\ Rev.\ Lett.\  {\bf 101}, 061101 (2008) \& J.~Abraham {\it et al.} [Pierre Auger Collaboration], Phys. Lett. B {\bf 685} (2010).

\bibitem{TAGZK}
  T.~Abu-Zayyad {\it et al.} [Telescope Array Collaboration],
  Astrophys.\ J.\  {\bf 768}, L1 (2013)
  [arXiv:1205.5067 [astro-ph.HE]].

\bibitem{g}
  K.~Greisen, Phys.\ Rev.\ Lett.\ {\bf 16}, 748  (1966)

\bibitem{zk}
  Z.~T.~Zatsepin and V.~A.~Kuz'min, Zh.\ Eksp.\ Teor.\ Fiz.\ Pis'ma Red. {\bf 4}, 144 (1966)

\bibitem{Gelmini:2005wu}
  G.~Gelmini, O.~E.~Kalashev and D.~V.~Semikoz,
  J.\ Exp.\ Theor.\ Phys.\  {\bf 106}, 1061 (2008)
  [astro-ph/0506128].

\bibitem{Aloisio:2015ega}
  R.~Aloisio, D.~Boncioli, A.~di Matteo, A.~F.~Grillo, S.~Petrera and F.~Salamida,
  JCAP {\bf 1510}, no. 10, 006 (2015)
  [arXiv:1505.04020 [astro-ph.HE]].

\bibitem{Saveliev}
  A.~Saveliev, L.~Maccione and G.~Sigl,
  JCAP {\bf 1103}, 046 (2011)
  [arXiv:1101.2903 [astro-ph.HE]].

\bibitem{Sokolov:2016lba}
  A.~V.~Sokolov and M.~S.~Pshirkov,
  arXiv:1611.04949 [hep-ph].

\bibitem{Gaisser:1993ix}
  T.~K.~Gaisser {\it et al.},
  Phys.\ Rev.\ D {\bf 47}, 1919 (1993).

   \bibitem{Abbasi2}
  R.~U.~Abbasi {\it et al.} [HiRes Collaboration], Phys.\ Rev.\ Lett.\  {\bf 104}, 161101 (2010) [arXiv:0910.4184 [astro-ph.HE]].

  \bibitem{Aab:2014aea}
  A.~Aab {\it et al.} [Pierre Auger Collaboration], Phys.\ Rev.\ D {\bf 90}, no. 12, 122006 (2014)
  [arXiv:1409.5083 [astro-ph.HE]].

  \bibitem{Hanlon}
  R.~U.~Abbasi {\it et al.} [Telescope Array Collaboration], 
  Astrophys.\ J.\  {\bf 858}, no. 2, 76 (2018) [arXiv:1801.09784 [astro-ph.HE]].

\bibitem{TA_AUGER_Composition_WG}
 V. De Souza et al.  Proceedings of the ICRC 2017, CRI167.

  \bibitem{Aab:2016enk} 
  A.~Aab {\it et al.} [Pierre Auger Collaboration],
  Phys.\ Rev.\ D {\bf 93}, no. 7, 072006 (2016) [arXiv:1604.00978 [astro-ph.HE]].

  \bibitem{Aab:2017cgk}
  A.~Aab {\it et al.} [Pierre Auger Collaboration], Phys.\ Rev.\ D {\bf 96}, no. 12, 122003 (2017) [arXiv:1710.07249 [astro-ph.HE]].

 \bibitem{Breiman}
  L.~Breiman et al., Wadsworth International Group (1984).

 
 \bibitem{Schapire}
  R.E.~Schapire, Mach.\ Learn.\ {\bf 5} (1990) 197.

  \bibitem{Krause}
  M.~Krause {\it et al.}, Astropart.\ Phys. {\bf 89}, 1 (2017) [arXiv:1701.06928 [astro-ph.IM]].

  \bibitem{Aab}
  A.~Aab {\it et al.} [Pierre Auger Collaboration], JCAP {\bf 1704} (2017) no.04, 009 [arXiv:1612.01517 [astro-ph.HE]].

  \bibitem{Abbasi}
  R.~Abbasi {\it et al.} [IceCube Collaboration], Phys.\ Rev.\ D {\bf 83} (2011) 012001 [arXiv:1010.3980 [astro-ph.HE]].

  \bibitem{Linsley}
  J.~Linsley, L.~Scarsi, Phys.\ Rev. {\bf 128} (1962) 2384.

  \bibitem{Teshima}
  M.~Teshima {\it et al.}, J.\ Phys.\ G {\bf 12}, 1097 (1986).

  \bibitem{Takeda}
  M.~Takeda {\it et al.}, Astropart.\ Phys.\  {\bf 19}, 447 (2003) [astro-ph/0209422].

  \bibitem{TAgammalim}
  T.~Abu-Zayyad {\it et al.} [Telescope Array Collaboration], Phys.\ Rev.\ D {\bf 88}, no. 11, 112005 (2013) [arXiv:1304.5614 [astro-ph.HE]].

  \bibitem{Takahashi:2011zzd} 
  Y.~Takahashi {\it et al.} [Telescope Array Collaboration],
  AIP Conf.\ Proc.\  {\bf 1367}, 157 (2011).

    \bibitem{Ros}
G.~Ros {\it et al.}, Astropart.\ Phys.\  {\bf 35}, 140 (2011) [arXiv:1104.3399 [astro-ph.HE]].

\bibitem{Heck}
D.~Heck {\it et al.}, Report FZKA-6019 (1998), Forschungszentrum Karlsruhe.

\bibitem{Ostapchenko}
  S.~Ostapchenko, Nucl.\ Phys.\ Proc.\ Suppl.\  {\bf 151}, 143 (2006) [hep-ph/0412332].

\bibitem{FLUKA}
  T.~T.~B\"ohlen {\it et al.},
 Nucl.\ Data Sheets {\bf 120}, 211 (2014)

 \bibitem{FLUKA2}
  A.~Ferrari, P.~R.~Sala, A.~Fasso and J.~Ranft, CERN-2005-010, SLAC-R-773, INFN-TC-05-11.

\bibitem{EGS4}
W.~R.~Nelson, H.~Hirayama, D.W.O.~Rogers,
SLAC-0265 (permanently updated since 1985).

 \bibitem{Hillas}
  A.~M.~Hillas, Nucl.\ Phys.\ Proc.\ Suppl.\  {\bf 52B}, 29 (1997).

  \bibitem{Kobal:2001jx}
  M.~Kobal [Pierre Auger Collaboration],
  Astropart.\ Phys.\  {\bf 15}, 259 (2001).

\bibitem{Gorbunov:2007vj}
  D.~S.~Gorbunov, G.~I.~Rubtsov and S.~V.~Troitsky,
  Phys.\ Rev.\ D {\bf 76}, 043004 (2007).

  \bibitem{Stokes}
  B.~T.~Stokes {\it et al.}, Astropart. Phys. {\bf 35}, 759 (2012).

 \bibitem{Agostinelli}
  S.~Agostinelli {\it et al.} [GEANT4 Collaboration], Nucl.\ Instrum.\ Meth.\ A {\bf 506}, 250 (2003).

\bibitem{TAdataMC}
   T.~Abu-Zayyad {\it et al.} [Telescope Array Collaboration],
  arXiv:1403.0644 [astro-ph.IM].

  \bibitem{AbuZayyad:2012ru} 
  T.~Abu-Zayyad {\it et al.} [Telescope Array Collaboration],
  Astrophys.\ J.\  {\bf 768}, L1 (2013) [arXiv:1205.5067 [astro-ph.HE]].

  \bibitem{Hocker}
  A.~Hocker {\it et al.}, PoS ACAT (2007) 040 [physics/0703039 [PHYSICS]].

  \bibitem{Freund}
  Y.~Freund, R.E.~Schapire, Proc. ICML (1996) 148.

\bibitem{ROOT}
  R.~Brun and F.~Rademakers, Proceedings AIHENP'96 Workshop, Lausanne, Sep. 1996, Nucl. Inst. \& Meth. in Phys. Res. A {\bf 389} (1997) 81-86, See also http://root.cern.ch/.

  \bibitem{TFractionFitter}
  R.~Barlow and C.~Beeston, Comp. Phys. Comm. {\bf 77} (1993) 219-228.


  \bibitem{Ostapchenko:2010vb} 
  S.~Ostapchenko,   Phys.\ Rev.\ D {\bf 83}, 014018 (2011) [arXiv:1010.1869 [hep-ph]].

\bibitem{Pierog}
 T.~Pierog, I.~Karpenko, J.~M.~Katzy, E.~Yatsenko and K.~Werner, Phys.\ Rev.\ C {\bf 92}, no. 3, 034906 (2015) [arXiv:1306.0121 [hep-ph]].

 \bibitem{Fletcher}
 R.~S.~Fletcher, T.~K.~Gaisser, P.~Lipari and T.~Stanev, Phys.\ Rev.\ D {\bf 50}, 5710 (1994).

 \bibitem{Aab:2016hkv}
 A.~Aab {\it et al.} [Pierre Auger Collaboration], Phys.\ Rev.\ Lett.\  {\bf 117} (2016) no.19,  192001, [arXiv:1610.08509 [hep-ex]].

 \bibitem{Takeishi}
R.~U.~Abbasi {\it et al.} [Telescope Array Collaboration], Phys.\ Rev.\ D {\bf 98}, no. 2, 022002 (2018)  [arXiv:1804.03877 [astro-ph.HE]].

   \bibitem{PierreAuger}
  P.~Abreu {\it et al.} [Pierre Auger Collaboration], Contributions to the 32nd International Cosmic Ray Conference, Beijing, China, August 2011 [arXiv:1107.4804 [astro-ph.HE]].

 \bibitem{Dedenko}
   A.~Sabourov {\it et al.}, Contributions to the 35th International Cosmic Ray Conference, Busan, South Korea, July 2017, PoS(ICRC2017)553.

    \bibitem{Aab:2017njo} 
  J.~Bellido {\it et al.} [Pierre Auger Collaboration],
  PoS ICRC {\bf 2017} 506, arXiv:1708.06592 [astro-ph.HE].

   \bibitem{Abraham}
 J.~Abraham {\it et al.} [Pierre Auger Collaboration], Phys.\ Rev.\ Lett.\  {\bf 100}, 211101 (2008) [arXiv:0712.1909 [astro-ph]].

  \bibitem{AbuZayyad3}
  T.~Abu-Zayyad {\it et al.} [Telescope Array Collaboration], ApJL \textbf{768} (2013) L1.

\end{thebibliography}
\end{document}